# A Survey of Pathways for Mechano-Electric Coupling in the Atria

Marta Varela[1,2*], Aditi Roy[2,3], Jack Lee[2]

[1] National Heart and Lung Institute, Faculty of Medicine, Imperial College London, London, UK
[2] Department of Biomedical Engineering, School of Biomedical Engineering & Imaging Sciences, King's College London, London, UK
[3] Department of Computing, University of Oxford, Oxford, UK

**\* Correspondence to:**
Dr Marta Varela
National Heart & Lung Institute
Imperial College London
4th floor, ICTEM Building
Hammersmith Hospital Campus
Du Cane Road, W12 0NN
London, UK
**e-mail:** marta.varela@imperial.ac.uk

## Competing Interests

## Keywords

## Abbreviations
| | |
|---|---|
| AF | Atrial fibrillation |
| AP | Action potential |
| APD | Action potential duration |
| BCL | Basic Cycle Length |
| $C_m$ | Transmembrane capacitance |
| CV | Conduction velocity |
| EP | Electrophysiology |
| ERP | Effective refractory period |
| MEC | Mechano-electric coupling |
| RMP | Resting membrane potential |
| SAC_NS | Stretch-activated current (non-specific) |
| SAC_K | Stretch-activated current (permeable to potassium ions) |

## Highlights
- Several pathways for mechano-electric feedback exist, including stretch-activated channels (non-specific and permeable to potassium ions), stretch-induced alterations in cell membrane capacitance and intra-cellular calcium handling, geometric effects and effects mediated by non-cardiomyocytes.
- In the atria, passive strains caused by ventricular contraction need to be considered when incorporating mechano-electro feedback in atrial electrophysiology models.

- We find that, for chronic stretch, stretch-induced capacitance changes dominate, leading to an overall increase in action potential duration and a reduction in conduction velocity, consistent with experimental studies.
- In the presence of passive stretch in atrial models, stretch-activated channels can mediate the appearance of delayed after-depolarisations and lead to rotor hypermeandering. These novel findings are likely to have implications for the genesis and maintenance of atrial arrhythmias.

## Abstract

Mechano-electric coupling (MEC) in atrial tissue has received sparse investigation to date, despite the well-known association between chronic atrial dilation and atrial fibrillation (AF). Of note, no fewer than six different mechanisms pertaining to stretch-activated channels, cellular capacitance and geometric effects have been identified in the literature as potential players. In this mini review, we briefly survey each of these pathways to MEC. We then perform computational simulations using single cell and tissue models in presence of various stretch regimes and MEC pathways. This allows us to assess the relative significance of each pathway in determining action potential duration, conduction velocity and rotor stability. For chronic atrial stretch, we find that stretch-induced alterations in membrane capacitance decrease conduction velocity and increase action potential duration, in agreement with experimental findings. In the presence of time-dependent passive atrial stretch, stretch-activated channels play the largest role, leading to after-depolarizations and rotor hypermeandering. These findings suggest that physiological atrial stretches, such as passive stretch during the atrial reservoir phase, may play an important part in the mechanisms of atrial arrhythmogenesis.

## 1. Introduction

Arrhythmias are one of the major causes of death worldwide, accounting for 17 million deaths each year (Srinivasan and Schilling, 2018). Despite being disturbances of the propagation of electrical signals in the heart, their devastating impact is caused by their perturbation of the heart's mechanical function. This relationship between electrical impulses and cardiac contraction is mediated by well-known electro-mechanical coupling pathways, involving intracellular calcium handling. It is also known that cardiac mechanics can influence the propagation of the electrical potential in the heart, although these mechano-electric coupling (MEC) mechanisms have received much less attention so far. Nevertheless, MEC effects can have a critical influence on cardiac function. The initiation of an arrhythmia through a high-impulse mechanical impact on the chest (*commotio cordis*) (Kohl et al., 1999) and the termination of arrhythmias using a similar mechanism (Pennington et al., 1970) are some of the most dramatic manifestations of MEC.

MEC is expected to also play an important role in atrial arrhythmias, particularly in atrial fibrillation (AF), the most common sustained arrhythmia. AF is characterised by a rapid, irregular and ultimately inefficient contraction of the atria. It affected 32.5 million people worldwide in 2010 and its incidence in Western Europe is expected to rise to 3% of all adults aged over 20 by 2030 (Kirchhof et al., 2016). AF is independently associated with a two-fold increase in all-cause mortality and is strongly associated with stroke, heart failure and cognitive impairment (Kirchhof et al., 2016). AF is thus often accompanied by substantial decreases in quality of life and a high rate of hospitalizations (Calkins et al., 2012) and, altogether, the economic burden of AF already amounts to 1% of total healthcare costs in the UK (Kirchhof et al., 2016).

The electrophysiological mechanisms underlying AF are notoriously complex. As a consequence, treatment to restore sinus rhythm to AF patients is unfortunately relatively inefficient. Catheter ablation, arguably the most successful form of rhythm control treatment, suffers from 3-year recurrence rates as high as 53% for a single procedure, which improve to 73% only following repeat procedures (Ganesan et al., 2013).

Chronic atrial stretch (i.e. atrial dilation) is considered both a predisposing factor for AF and a consequence of AF-induced structural remodelling (Schotten et al., 2003) and, as such, has been used to predict the success of catheter ablations in AF (Berruezo et al., 2007; Varela et al., 2017). Animal and human studies of chronic atrial stretch widely demonstrate an enhanced sensitivity to AF induction and (often heterogeneous) conduction slowing but variable or negligible effects on the effective refractory period (ERP), as recently reviewed by (Thanigaimani et al., 2017). The mechanisms through which atrial dilation may promote AF are complex and may involve activation of multi-organ signalling pathways and a concomitant development of atrial fibrosis (Schotten et al., 2003) and inflammation (Verheule et al., 2004). Other direct pathways may involve an increase in the atrial surface area available for re-entrant electrical circuits and enhanced conductance of stretch-activated channels experienced by hypertrophied cardiomyocytes (Kamkin et al., 2000).

Acute atrial stretch is also known to lead to a plethora of electrophysiological changes, although many studies offer contradictory findings, particularly in relation to stretch-induced changes in ERP (Ravelli, 2003). Most animal and human studies suggest that acute atrial stretch leads to a reduction in overall conduction velocity (CV), an increase in CV heterogeneity and increased susceptibility to AF (Coronel et al., 2010; Ravelli et al., 2011; Thanigaimani et al., 2017; Walters et al., 2014). Stretch-induced after-depolarizations and ectopic beats have also been reported (Franz, 2000).

Despite all the evidence for the important role MEC can play in atrial arrhythmias, there have been few computational modelling studies investigating the pathways through which MEC can contribute to atrial arrhythmias. When included, only certain MEC pathways have been considered in each study (Kuijpers et al., 2011, 2007; Satriano et al., 2018) making it difficult to assess which of the MEC contributions are likely to play the most significant role in the genesis and maintenance of AF. Furthermore, most of the methodology for incorporating MEC in atrial models borrows heavily from ventricular studies, not making allowance for important effects such as the large passive strain experienced by the atria during ventricular contraction (Hoit, 2014). In this article, we aim to catalogue the proposed MEC pathways and perform initial simulations to assess their potential role in atrial arrhythmogenesis. We expect this to contribute to the creation of a set of standard mechanisms for incorporating MEC in computational simulations of atrial electrophysiology.

## 1.1. Computational Modelling of Mechano-Electric Coupling

In this section, we briefly revisit the current standard methods for incorporating MEC in EP simulations, which have, in general, been created to study mechano-electric effects in the ventricles. We then discuss the alterations to these models that may make them more suitable for modelling MEC in the atria, before introducing the MEC pathways under study.

To date, computational models of atrial arrhythmias usually restrict themselves to electrophysiology, disregarding any changes to atrial EP properties that may arise from the cardiac mechanical function. An exception is the recent study by (Satriano et al., 2018), which models strongly-coupled electrical and mechanical left atrial function in physiological conditions, including some MEC pathways ($K^+$-permeable stretch-activated channels). EP studies that include MEC typically involve the following steps:

1. Modelling the propagation of the action potential (AP) in tissue using an existing electrophysiological (EP) model;

2. Introducing forward electro-mechanical coupling by computing the active tension, $T_a$, generated by local changes in transmembrane voltage, $V_m$. This can be accomplished in two different formalisms:

- a direct phenomenological relationship, usually an ordinary differential equation where changes in $T_a$ depend explicitly on $V_m$ (Nash and Panfilov, 2004; Panfilov et al., 2007). This is the usual approach employed when using a phenomenological EP model with a reduced number of channels, such as the Fenton-Karma (Fenton and Karma, 1998) or the Aliev-Panfilov (Aliev and Panfilov, 1996) EP models.
- a detailed physiological description of myofilament physiology, where intracellular calcium and its binding to sarcolemmal buffers mediate the relationship between $T_a$ and $V_m$ (Land et al., 2017; Rice et al., 2008). This approach is typically parameterised using ventricular data and is usually employed when the EP model is very detailed and already involves explicit calculations of intracellular calcium concentration and calcium buffer dynamics.

3. Calculating the deformation (strains) experienced by the heart in response to the active tension. This represents solving a classical solid mechanics problem, which relies on assumptions about the mechanical properties of myocardium. Most studies treat the myocardium as a hyperelastic medium. A detailed review of the constitutive models used to describe the passive mechanical properties of the myocardium can be found in (Avazmohammadi et al., 2019).

Although the template above can be used as a starting point for modelling MEC in the atria, important differences between atrial and ventricular function should be taken into account:

- Most of the strain experienced by the atria is passive (Fig 1A). This is caused by both ventricular contraction and passive filling of the atria with blood (reservoir phase; systolic strain $\varepsilon_s$ = 29-48%) and subsequently by emptying of atrial blood to the ventricles (conduit phase; early diastolic strain $\varepsilon_e$ = 9-27%). Echocardiographic and CINE MRI studies suggest that these passive atrial strains greatly exceed those experienced during active atrial contraction (booster pump; active strain $\varepsilon_a$ = 8-20%). All quoted LA strain values are adapted from the review by (Hoit, 2017) and are given as absolute values, with the atria at its smallest size (just before the QRS complex) as the reference configuration. $\varepsilon_s$ corresponds to atrial stretch, whereas $\varepsilon_e$ and $\varepsilon_a$ are a measure of atrial contraction.

    These measurements are typically performed by measuring changes in the length of a contour on a 2D atrial image and therefore the measured strain values do not represent length changes across the direction of maximum stretch/contraction, i.e. principal strain values. They are furthermore measured across large segments of the atria and therefore likely not to be representative of true regional strains. Finally, differences in imaging technique and methodology lead to a large spread in measured strain values (Tobon-Gomez et al., 2013).

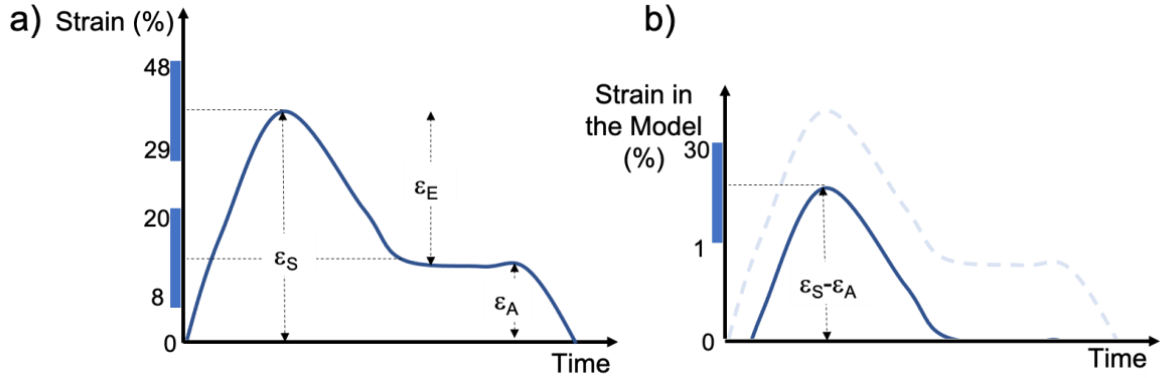

Figure 1: Left atrial strain as a function of time. (Time 0 corresponds to the QRS complex in an electrocardiogram.) A) Global left atrial strain curve, as measured using CINE MRI and echocardiography. Data from (Hoit, 2017). B) Positive left atrial strain inputted into computational models (see below).

- To date, few studies have characterised the mechanical properties of the atria and, when this is has been done, a large variability in properties has been found across subjects and atrial regions (Bellini et al., 2013). It is also not clear what the appropriate boundary conditions in atrial mechanical models are (Di Martino et al., 2011). Atrial-specific constitutive models have not yet been adopted in most studies, except for the model proposed in (Bellini et al., 2013).

### 1.2. Pathways for Mechano-Electric Coupling in the Atria

In this section, we survey the pathways for MEC that are likely to play a role in the atria, as show in the diagram in Fig 2. These include mechanically gated channels (such as non-specific and $K^+$-permeable channels); alterations in cell capacitance and conductivity mediated by mechanical effects, as well as changes in the dimensions and geometry of the propagation medium induced by mechanical deformations. Effects on fibroblasts and other non-muscular cell types present in myocardium and on the intracellular mechanisms of electro-mechanical coupling have also been proposed.

The source of experimental data for these pathways is heterogeneous and often comes from ventricular cells or models. When relevant for the simulations we carried out, we provide expressions relating the considered pathways and the local stretch, $\lambda$. $\lambda$ represents the fractional stretch experienced by local myocytes – more rigorously, $\lambda = \sqrt{\det(C)}$ where C is the right Cauchy-Green deformation tensor and $\lambda = 1$ corresponds to the reference, no-stretch configuration. $\lambda$ is not directly related to the global strains represented by $\varepsilon$ in Fig 1A. Given the current lack of measurements of $\lambda$ in the atria, we treat $\varepsilon_S$-$\varepsilon_A$ (Fig 1B) as a suitable estimate for the largest local stretch, $\lambda$, experienced by atrial myocytes.

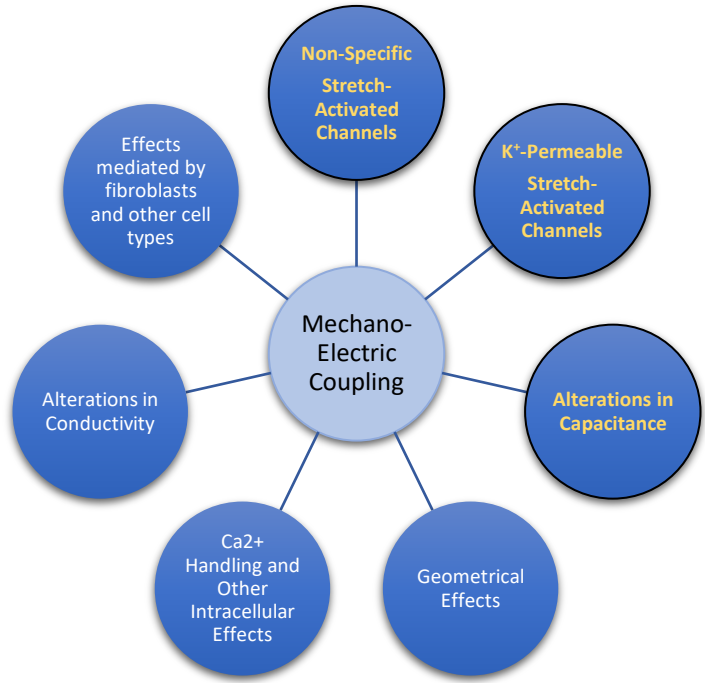

Figure 2: Mechano-electric coupling pathways considered in this study. The effects on atrial EP of the pathways in yellow font are investigated in the current study using computational simulations.

### 1.2.1 Mechanically gated channels

Several mechanically gated channels have been identified in the human heart. These include stretch-activated channels (SACs) and volume-activated channels as well as putative channels activated by changes in membrane curvature or thickness. Channels whose conductance depends on hydrostatic pressure have also recently been characterised in immune cells (Solis et al., 2019) and leukaemia cell lines (Hui et al., 2014). If present in cardiac tissue, these may also play an important role in MEC.

In addition to all these, it is known that the properties of several ionic currents, such as $I_{Na}$, $I_{CaL}$, $I_{KATP}$ and $I_{K1}$, can be affected by mechanical stimuli, as detailed in a recent review (Peyronnet et al., 2016). Here, we will focus on stretch-activated channels, which are arguably the most important mediators of MEC at the ionic channel-level.

Stretch-activated channels (SACs) are protein complexes that span the sarcolemma and whose open probability increases in the presence of local membrane stretch. SACs can be divided into two broad categories: K$^+$-permeable SACs (SAC_K) and non-specific SACs (SAC_NS).

As with other potassium currents, the reversal potential for SAC_K is slightly more negative than the resting membrane potential (RMP). SAC_K thus contributes to the hyperpolarization/repolarization of the cell and, when activated, can shorten the action potential duration (APD) and the effective refractory period (ERP) and make RMP more negative (Peyronnet et al., 2016). There are several molecular candidates for SAC_K, such as TREK-1, TREK-2 and TRAAK. We model SAC_K as proposed by (Healy and McCulloch, 2005):

$$I_{SAC\_K} = \frac{g_{SAC_K}}{1+exp^{\frac{1.185-u}{1.316}}}(\lambda - 1) \qquad [1]$$

where u is the transmembrane potential scaled to the [0, 1] interval, as is typically done in low-dimensional ionic models such as the Fenton-Karma model (Fenton and Karma, 1998). $g_{SAC\_K}$, the channel conductance, has a value of 0.024 μA as the mean of the previously proposed interval (0.01 to 0.04 μA) (Healy and McCulloch, 2005).

SAC_NS, on the other hand, are permeable to a wider range of ions, placing their reversal potential between –20 and 0 mV (Peyronnet et al., 2016). If active during the repolarisation phases of the action potential, SAC_NS can increase APD and ERP and also make the RMP less negative. Piezo 1 or 2 and TRP channels are the most likely candidates for SAC_NS's molecular counterpart. SAC_NS are modelled as a passive current (Zeng et al., 2000):

$$I_{SAC\_NS} = g_{SAC\_NS}(u - u_{rev})(\lambda - 1) \qquad [2]$$

We use $g_{SAC\_NS}$ = 0.025 μA and $u_{rev}$ = 0.837 (corresponding to -10 mV), as proposed in the literature (Timmermann et al., 2017).

Models of SAC_K and SAC_NS typically assume that these channels are either homogeneously distributed across the heart or that there is a ventricular transmural gradient (Healy and McCulloch, 2005). It is furthermore usually assumed that the channels' response does not depend on whether the sensed stretch is longitudinal or transverse relative to the myofibres. The response of these channels to an external stretch is also assumed to be instantaneous. Few experimental studies have specifically focused on atrial cells – an exception is (Kamkin et al., 2003).

To date, SACs have been the most studied MEC mechanism using computational simulations. As explained above, most studies involve adding formulations of stretch-activated currents to an established electrophysiology model, which can be detailed (e.g. (Ten Tusscher et al., 2003)) or phenomenological e.g. (Fenton and Karma, 1998)), coupled with a model for active tension generation (Brocklehurst et al., 2017; Colli Franzone et al., 2017; Hu et al., 2013; Keldermann et al., 2009; Nash and Panfilov, 2004; Panfilov et al., 2007; Satriano et al., 2018; Weise and Panfilov, 2019). Findings from these studies corroborate the role of MEC in influencing spiral wave meandering (Brocklehurst et al., 2017; Colli Franzone et al., 2017; Dierckx et al., 2015; Radszuweit et al., 2015) and, in some conditions, spiral wave breakup (Keldermann et al., 2010; Panfilov et al., 2007; Weise and Panfilov, 2017) and even spiral wave initiation (Weise and Panfilov, 2011; Yapari et al., 2014).

Whereas most of these studies used ventricular geometries and/or ventricular ionic models, some have focused on the atria, including atrial ionic models that include AF remodelling (Brocklehurst et al., 2017) and atrial geometries (Kuijpers et al., 2011, 2007). Other atrial studies have included externally-imposed stretches (although not designed to mimic the ones experienced by the atria) and found that these affected the anchoring of spiral waves, often in the presence of other arrhythmogenic effects (Yamazaki et al., 2012, 2009). Recently proposed atrial electromechanical models (Augustin et al., 2019; Moyer et al., 2015) have typically not included SACs or other MEC pathways. A notable exception is the study by Satriano and colleagues (Satriano et al., 2018), who considered the effect of stretch-activated channels on left atrial mechanics, but not on arrhythmogenesis.

### 1.2.2 Atrial myocyte capacitance

Some studies have shown a dependence of the capacitance of the cardiomyocyte cell membrane on local stretch (Mills et al., 2008). Although the sparsity of experimental data leads to ambiguities in the optimal choice of model, we follow here the Hill function formalism previously introduced for ventricular cells (Oliveira et al., 2015):

$$C_m = 1 + \frac{(\lambda-1)^6}{0.04^6 + (\lambda-1)^6} \qquad [3]$$

Here, we treat the myocyte capacitance as a relative variable, with $C_m$ = 1 µA s/m$^2$ in the absence of stretch ($\lambda$ = 1).

### 1.2.3. Ca$^{2+}$ handling and other intracellular processes

It is increasingly recognised that stretch can also affect intracellular processes, including the release of bound calcium from troponin C in myofilaments (Schönleitner et al., 2017; ter Keurs et al., 1998) and an increase in the open probability of ryanodine receptors, through the so-called X-ROS signalling processes (Prosser et al., 2011). These effects can lead to changes in electrophysiology by affecting Ca$^{2+}$ handling within myocytes, with some authors even proposing that they may be responsible for ectopic beats (ter Keurs et al., 1998; Timmermann et al., 2019). Moreover, alterations in intracellular Ca$^{2+}$ induced by stretch may also affect the conductance of ionic channels sensitive to intracellular Ca$^{2+}$, as discussed in more detail in (Calaghan et al., 2003).

These Ca$^{2+}$-mediated mechanisms are not amenable to investigations using simplified models of electrophysiology and will therefore not be explored using computational simulations in the current study. There are reviewed in more detail in (Timmermann et al., 2017). We also do not consider here the interesting effects stretch may have on gene expression in cardiomyocytes and other cellular constituents of the myocardium. A detailed review of these effects can be found in (Saucerman et al., 2019).

### 1.2.4. Effect of stretch on non-cardiomyocytes

As arrhythmogenic mechanisms are unveiled, the role of cardiac cells other than atrial cardiomyocytes has become increasingly prominent. Atrial fibrotic remodelling is believed to be of particular importance, as fibrotic regions are known to be sites of slow and/or discontinuous conduction velocity (Nguyen et al., 2014) and likely rotor anchoring sites (Roy et al., 2018). Less is known about how fibrosis may play a role in arrhythmia mechanisms through MEC. Its high stiffness relative to healthy myocardium suggests that cardiomyocytes surrounding areas of fibrosis may undergo comparatively smaller deformations than areas of healthy myocardium further afield. This could further contribute to heterogeneities in EP properties across the atria, promoting arrhythmias (Weise and Panfilov, 2012). Further characterisation of the electrical and mechanical properties of non-cardiomyocytes is needed to study these effects in more detail.

### 1.2.5. Geometric effects and effects on conductivity

In some studies, the change in pathlength of the trajectory of the action potential induced by stretch has also been incorporated, as suggested by among others (Colli Franzone et al., 2017; Nash and Panfilov, 2004; Panfilov et al., 2007; Quarteroni et al., 2017). In practical terms, this corresponds to solving the diffusion PDE governing the propagation of the action potential in a deforming geometry rather than in undeformed (material) coordinates. This approximation assumes that conduction velocity is dominated by intracellular conductivity, disregarding conduction delays at (intercellular) gap junctions. However, experimental studies suggest that conduction velocity across the cytoplasm of cardiomyocytes is several orders of magnitude higher than CV across gap junctions in 1D tissue strands, making conduction of the electrical signal almost saltatory (Rohr, 2004). This supports the suggestion that the geometric implications of the stretch of the cytoplasm of cardiomyocytes are negligible (Pfeiffer et al., 2013).

The conduction velocity in cardiomyocytes embedded in 3D tissue, nevertheless, appears to depend more on cytoplasmic conduction, presumably due to the averaging effects of lateral gap junctions (Rohr, 2004). This implies that modelling the geometric effect of stretch on conduction velocity may be more complex than what has hitherto been considered.

In addition to the above, it is not clear whether electrical conductivity (typically incorporated in the electrical diffusion tensor, D, in modelling studies) is directly affected by stretch. Some studies have suggested that intercellular connectivity at gap junctions may be increased (Zhuang et al., 2000) or decreased (Kamkin et al., 2005; Mills et al., 2008) by stretch, whereas other have proposed that intracellular conductivity increases with (moderate) strains (McNary et al., 2008) or is affected by mechanical stress (Loppini et al., 2018). Other studies have instead modelled the geometric effect of stretch by assuming a reduction in the number of gap junctions per unit length of chronically dilated atrial myocardium (Kuijpers et al., 2007) or alterations in the space constant of the tissue (Oliveira et al., 2015).

## 2. Methods

To further investigate which of the above MEC ways may have the largest contribution for atrial arrhythmogenesis, we conducted some computational simulations. Of the above pathways, we focused on stretch-activated channels and stretch-induced changes in membrane capacitance, as we found these were some of the best candidates for easy inclusion in future atrial EP simulations: the selected pathways are comparatively well-characterised experimentally and amenable to being included in EP simulations with only minor alterations. We therefore did not explicitly model the more complex stretch-induced geometric effects in this study.

Computational simulations were carried out in 3 different scenarios. First of all, we modelled MEC pathways in single cells to gain an understanding of the impact of the proposed pathways on action potential morphology and properties such as action potential duration (APD) and resting membrane potential (RMP). We also modelled the same MEC pathways in a cuboid-shaped continuum of atrial cells in isotropic and anisotropic conditions when activated by a planar wave, as in sinus rhythm. This allowed us to determine stretch's impact on conduction velocity and on AP properties in the presence of electrotonic coupling. Finally, we investigated, in the same cuboid geometry and in isotropic conditions, the effect of MEC on the behaviour and dynamics of the re-entrant circuits (rotors) likely to be responsible for AF.

We carried all simulations under two distinct stretch protocols: 1. a time-independent stretch of constant amplitude, in order to model chronic atrial stretch or 2. a time-dependent stretch with an amplitude similar to Fig 1B, to model the passive stretch experienced by the atria in physiological conditions. For simplicity, we assumed that all cells in the cuboid tissue experienced the same (homogeneous) stretch at each point in time, even in the presence of rotors.

### 2.1 Single Cell Simulations

To study the effect of MEC on EP properties, we solved, in a single cell, the 3-variable Fenton Karma model (Fenton and Karma, 1998) parameterised to approximate the action potential from AF-remodelled atrial cells (Goodman et al., 2005; Roy et al., 2018). We scaled V, the transmembrane potential, such that V=0 corresponds to -80 mV and V=1, to 3.6 mV, as in the remodelled atrial Fenton-Karma model proposed by Goodman (Goodman et al., 2005). We used a forward Euler scheme ($\Delta t$ = 5 $\mu$s) implemented in Matlab, with a basic cycle length (BCL) of 400 ms to model rapid atrial activation.

All single-cell results are shown after a transition period of 12 cycle lengths. We incorporated each of the following MEC pathways in the model:

a) K$^+$ permeable stretch activated channels (eq 1)
b) Non-specific stretch activated channels (eq 2)
c) Stretch-induced changes in membrane capacitance (eq 3)
d) Simultaneous presence of effects a-c.

In the presence of:

1. No stretch;
2. Constant stretch of magnitude 1 to 3%;
3. Time-variable stretch with a peak of 30% whose time course mimics the global left atrial strains measured using echocardiography (see Fig 1B), with data taken from (Montserrat et al., 2015). Negative stretches (contractions) were not taken into account, as in previous studies (Panfilov et al., 2007; Timmermann et al., 2017). These measured strains were synchronised with the action potential, such that the active contraction of the LA ($\varepsilon_A$ in Fig 1A) took place 10 ms after the upstroke of the action potential for every cell.

The conditions imposed in 2 aimed to model a chronic stretch situation, whereas 3 sought to model the passive stretch experienced by the atria during the normal reservoir phase. We chose the stretch values for condition 2 to match those used in the experiments from which the used MEC pathway formalisms were derived. To allow direct comparisons between conditions 2 and 3, we additionally ran simulations with constant stretch values of 30% in condition 2.

To study how resting membrane potential and 90% action potential duration depend on the rate at which the cells were excited, we calculated APD$_{90}$ restitution curves for all MEC pathways (a-d) in the presence of constant stretch.

## 2.2 Tissue Simulations

We additionally solved the monodomain equation $\frac{\partial V}{\partial t} = \vec{\nabla}.(D\vec{\nabla}V) - \frac{I_{ion}}{C_m}$ on a 60 x 60 x 1.8 mm$^3$ cuboid, using the same remodelled atrial Fenton-Karma model to model the ionic currents I$_{ion}$. In this formalism, the transmembrane potential, V, at a given location can vary in time through the action of I$_{ion}$ or through the diffusion of V to/from neighbouring cells with a driving force parameterised by the diffusion tensor D.

We used an in-house MPI C solver, with centred finite differences ($\Delta$x = 0.3 mm) and a forward Euler scheme ($\Delta$t = 5 µs). Neumann boundary conditions were implemented in all simulations. Action potentials were initiated by raising the transmembrane potential in the left-most 5 cells (corresponding to 1.5 mm) at the boundary of the tissue at a basic cycle length of 400 ms. This aimed to model atrial activation in sinus rhythm in tachycardic conditions.

Simulations were performed for all combinations of effects a-d and stretch protocols 1-3 sensed uniformly by the entire tissue. Simulations were carried out in isotropic conditions with a diffusion coefficient of 0.1 mm$^2$/ms, corresponding to a baseline mean conduction velocity of approximately 60 cm/s, typical of AF remodelled atria (Varela et al., 2016). We also performed simulations in anisotropic conditions with D$_{longitudinal}$ = 0.3 mm$^2$/ms and D$_{transverse}$ = 0.03 mm$^2$/ms, with fibres aligned with y-axis (orthogonal to the direction of propagation of the action potential).

Conduction velocity and APD$_{90}$ were computed across the tissue, as detailed in (Varela et al., 2013). We additionally initiated a rotor using a cross-field protocol in the centre of the cuboid tissue, in the isotropic conditions described above. The position of the rotor tip was tracked for 2s, as described in (Roy et al., 2018).

## 3. Results

### 3.1 Single Cell Simulations

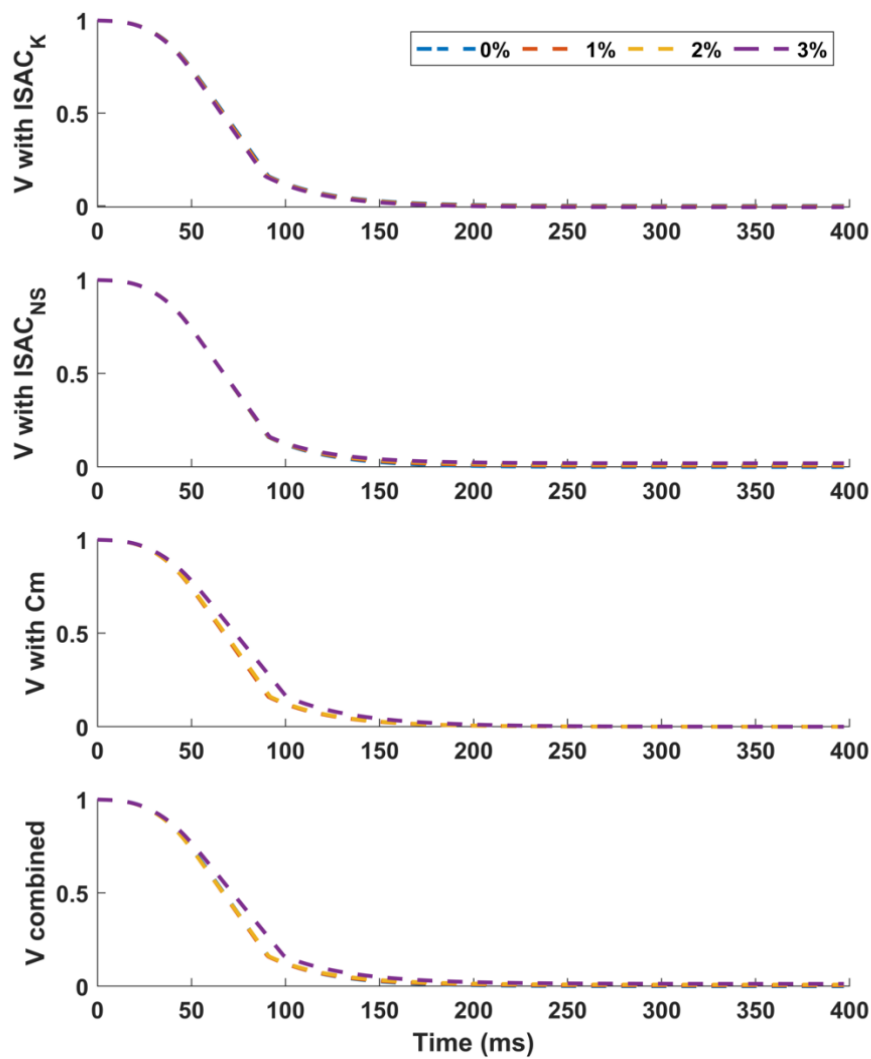

Figure 3: Transmembrane potential, V, under the effect of constant stretch ranging from 1-3% and considering effects of (from top to bottom) of the following MEC pathways: a) ISAC_K, b) ISAC_NS, c) capacitance changes and d) simultaneous effect of pathways a)-c).

The presence of constant stretches of up to 3% led to small changes in AP morphology (Fig 3), with observable changes only under the effect of Cm, which broadened the action potential and increased the RMP. As shown in Fig 4, SAC_K had a negligible effect on APD$_{90}$ (Fig 3A) whereas ISAC_NS tended to shorten it (Fig 3B) and stretch-induced alterations in cell capacitance, to increase it (Fig 3C). When combining all three effects, APD$_{90}$ was changed by less than 2% (Fig 3D and Fig 6A). The observed changes were qualitatively similar for different pacing cycle lengths ranging from 100 to 800 ms (Fig 4).

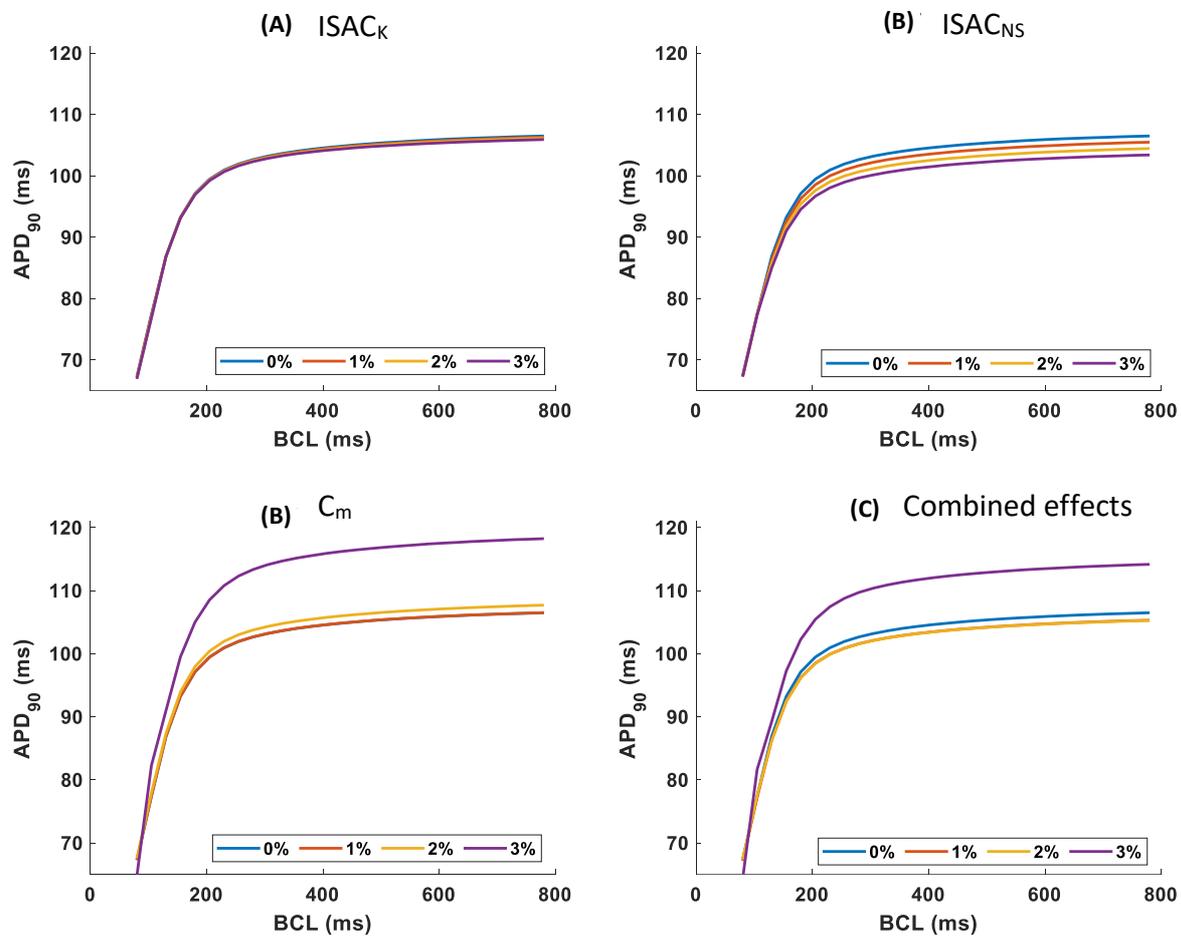

Figure 4: APD$_{90}$ restitution curves under constant stretch for all considered MEC pathways.

Under the action of a time-dependent passive stretch with a peak of 30%, AP morphology was altered, showing after-depolarizations (see Fig 5b). These were not observed under the presence of constant stretches of similar amplitude (Fig 5b). Fig 5c shows that SAC_NS is the main contributor to the appearance of these after depolarizations. SAC_K, on the other hand, has the opposite effect to ISAC_NS in these conditions, causing a repolarization dip in the action potential. For the current parameterisation of the SAC channels, the combined effect of these MEC pathways gives rise to an afterdepolarization. C$_m$ does not play a role in appearance of these changes in V after the action potential.

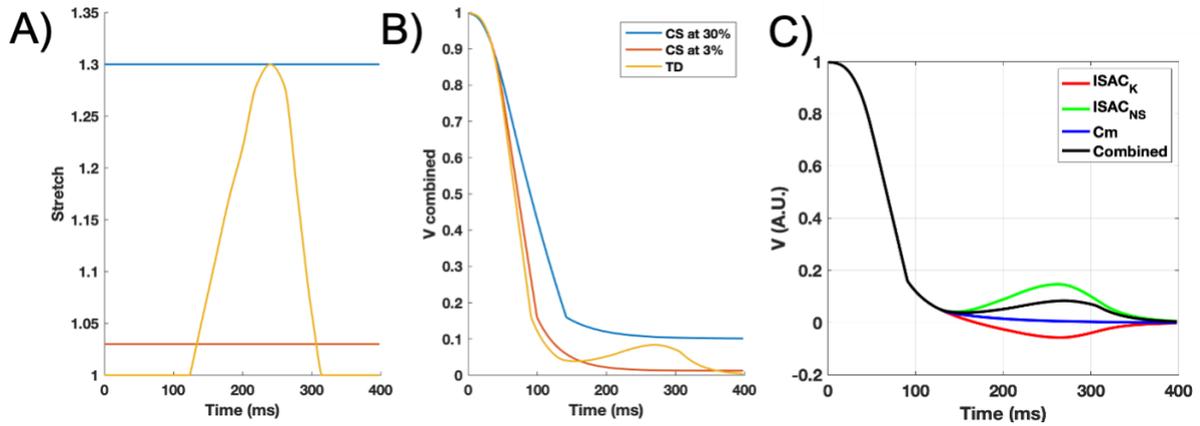

Figure 5: a) Stretch protocols corresponding to the transmembrane potential, V, curves of the same colour on panel b), ie. blue and red: constant stretch (CS) at different values and yellow: time dependent (TD) stretch following the curve displayed on the right. C) V under time-dependent stretch for all the 4 pathways considered.

## 3.2 Tissue Simulations

Tissue-level simulations confirmed the APD trends seen in single-cell simulations, with modest APD increases driven mostly by stretch-induced capacitive changes, as shown in Fig 6a. The inclusion of SAC_NS led to a marginal increase in CV, which was dwarfed by the reduction in CV (10% for a 3% uniform permanent stretch) induced by stretch modulation of Cm – see Fig 6b. Similar relative changes in $APD_{90}$ and CV were observed in isotropic and anisotropic conditions.

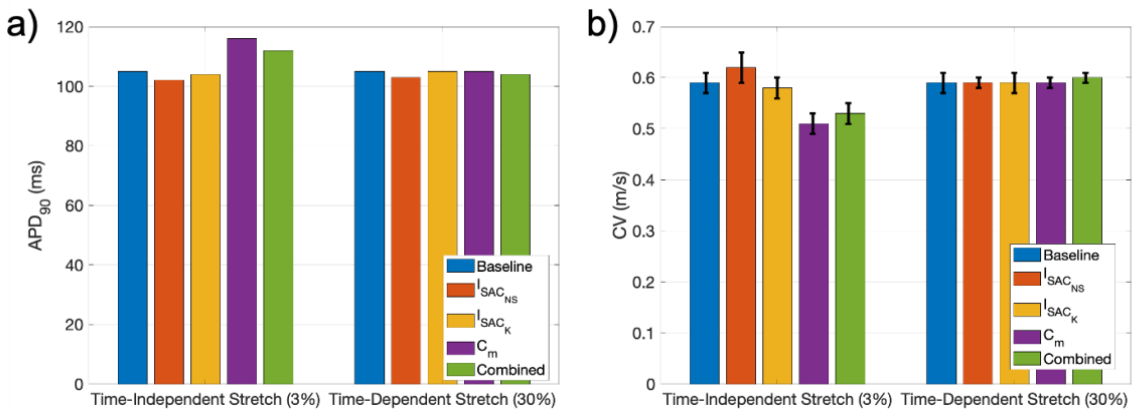

Figure 6: APD90 and CV for time-independent and time-dependent stretches in baseline conditions and under the action of the 4 considered MEC pathways.

Simulated rotors traced the stable epitrochoidal meandering pattern characteristic of this model in both baseline conditions and in the presence of uniform constant stretch – see Figure 7a. In the presence of the time-dependent stretch and the associated after-depolarizations, however, the hypermeandering greatly increased and the rotors terminated early by interacting with tissue boundaries in our model (Figure 7b).

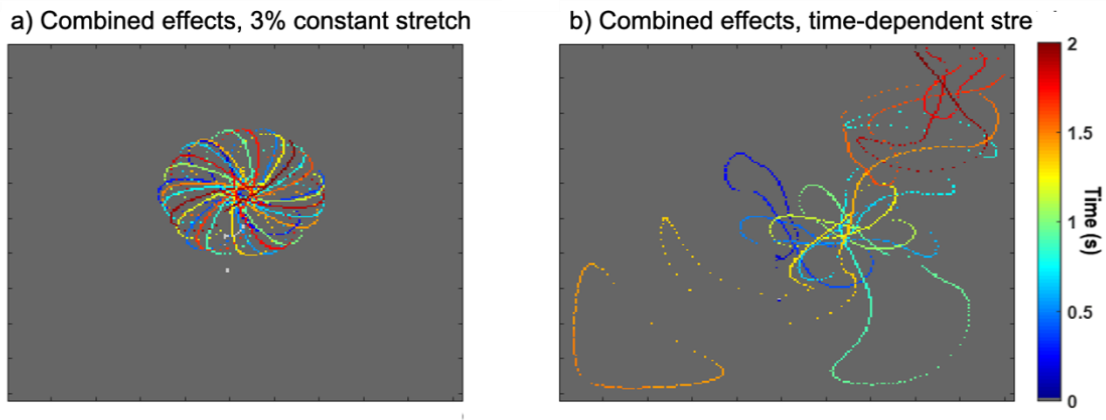

Figure 7: Trajectory of the tip of rotors initiated using a cross-field protocol. a) In the presence of constant 3% stretch, the rotor tip follows an epitrochoidal trajectory, similar to what is observed in the baseline (no stretch) case. b) In the presence of time-dependent stretch (30% stretch at peak), the rotor tip hypermeanders and quickly reaches a tissue boundary where it terminates. The trajectory of the rotor is colour-coded according to the time since the rotor initiation.

## 4. Discussion

In this manuscript, we surveyed the main mechanisms responsible for mechano-electric coupling in the atria. We stress, in particular, how the atria are subjected to large physiological passive external strains (from haemodynamic filling and ventricular systole), which have typically not been considered when studying MEC in these chambers. The list of potential pathways for MEC in atrial tissue is extensive (see Fig 2) and complex, making detailed experimental characterisations of each of these pathways very challenging. This is especially true for atrial tissue, as most studies are carried out in ventricular myocytes. Furthermore, it is not clear how some MEC pathways, especially those concerning to stretch-induced changes in conductivity and tissue size, should be implemented in computational EP models.

Despite these challenges, we performed computational simulations of three reasonably well-characterised effects that are easy to implement in EP simulations: non-specific and $K^+$-permeable stretch-activated current and stretch-induced alterations in $C_m$.

When found that in the presence of chronic stretch (modelled as a constant stretch of amplitude <3%), atrial action potentials remained relatively unchanged, with small increases in APD duration and RMP caused predominantly by capacitance-mediated effects. These effects were also observed in simulations performed in cuboidal atrial tissue, where CV was also reduced (as observed in clinical studies (Ravelli, 2003; Schotten et al., 2003)), also through the action of stretch-induced capacitance changes. Rotor dynamics were unaltered in these chronic stretch conditions.

In our simulations, chronic atrial stretch thus affects global atrial EP properties mostly though stretch-induced alterations in membrane capacitance. The reduction in CV is expected to be pro-arrhythmic by effectively reducing the size of re-entrant circuits and thus allowing a higher number of these to be sustained in the atria. For the same reasons, the observed lengthening of $APD_{90}$ can increase the effective rotor size and is, in principle anti-arrhythmic. Increases in RMP, on the other hand, may make the cells more excitable and thus more susceptible to ectopic firing.

When taking into account time-dependent passive stretches (typically experienced by atrial cells after the end of the refractory period), there was little difference in EP metrics such as APD or CV. An imbalance between ISAC_NS and ISAC_K, however, led to the appearance of after-depolarizations in single cell models. These are expected to be pro-arrhythmic by facilitating spontaneous depolarizations of the cell. Furthermore, these after-depolarizations lead to perturbations in the waveback of induced rotors, breaking up its typical epitrochoidal rotation pattern and causing the rotor to hypermeander, increasing the tissue area it occupies. Although in our simple cuboid tissue model, this hypermeandering led to an early termination of the rotor through interactions with tissue boundaries, in realistic atrial geometries the observed hypermeandering may prevent the rotor from anchoring to well defined structures such as the pulmonary veins or atrial fibrotic patches and thus become more difficult to ablate.

The observed after-depolarizations come from a balance between the two stretch-activated currents, whose parameterization is subjected to some uncertainty. We note that the afterdepolarizations are likely to take a different form for different parameterisations of ISAC_K and ISAC_NS which are compatible with experimental data. It is additionally not clear whether these after-depolarizations are relevant in atrial tissue, which typically does not undergo realistic haemodynamic stretches in experimental EP studies. In the atria, stretch is also expected to be heterogeneous, whereas we modelled homogeneous stretches across the entire tissue. If relevant, these MEC effects are likely to be particularly significant in the presence of asynchronous atrial activation, such as in AF, where they may contribute to heterogeneity in EP properties.

Future simulations conducted in realistic atrial geometries will study the implication of these effects to atrial arrhythmias in greater depth. Similarly, inhomogeneous stretches (passive or due to active atrial contraction), the presence of heterogeneous fibrosis or gradients in mechanical or EP properties across the atria will be incorporated in future studies.

Progress in studying the contributions of MEC to atrial EP, especially to atrial arrhythmias, is dependent on more detailed atrial-specific experimental characterization of the different MEC pathways. This will enable the parallel creation of more realistic computational models which include MEC effects and can shed light into the role of MEC in atrial fibrillation mechanisms. In particular, it is not clear how some of the surveyed MEC pathways (e.g. capacitance alterations under stretch) may depend on the direction of the applied stretch. This will be essential for future adequate modelling of MEC in realistic atrial geometries, which include the anisotropy induced by the complex atrial fibre orientations and region-specific strains.

## 5. Conclusions

In this paper, we discuss the many pathways through which MEC can affect atrial EP. We stress that the stretches that atria are subjected to are substantially different from those experienced by the ventricles and that this should be taken into account in atrial EP models than include MEC. In our simulations, we found that accounting for passive atrial stretches similar to those observed in healthy atria led to after-depolarisations and enhanced rotor meandering. We also observed reductions in CV and increases in $APD_{90}$ in the presence of chronic atrial stretch, consistent with experimental studies. These findings suggest that MEC effects are likely to play an important role in the atria. Further experimental characterization of MEC pathways and modelling studies are required to better understand MEC's role in atrial arrhythmias such as AF.

## Acknowledgements
We thank Mr Charles Houston and Dr Rasheda Chowdhury for assistance with the literature review.

This work was supported by the British Heart Foundation [RE/18/4/34215]; Wellcome/EPSRC Centre for Medical Engineering [RE/18/4/34215] and the EPSRC Medical Image Analysis Network [EP/N026993/1].